\documentclass[12pt]{article}
\usepackage{epsf}
\usepackage{cite}
\begin{document}
\begin{center}
{\bfseries SUPERNARROW DIBARYONS AND EXOTIC BARYONS WITH SMALL MASSES}

\vskip 5mm

L.V. Fil'kov
\vskip 5mm

{\small{\it
Lebedev Physical Institute, Moscow, Russia} \\
{\it E-mail: filkov@sci.lebedev.ru}}
\end{center}

\vskip 5mm

\begin{center}
\begin{minipage}{150mm}

\centerline{\bf Abstract}

A search for supernarrow dibaryons (SND) in $pd$ interactions is reviewed.
Narrow peaks at masses 1904, 1926, and 1942 MeV have been observed in the
missing mass $M_{pX_1}$ spectra of the reaction $pd\to p+pX_1$. The
analysis of the data obtained leads to the conclusion that these peaks are
most likely SNDs. The possible interpretation of the peaks, found in the
$M_{X_1}$ mass spectra at 966, 986, and 1003 MeV and in the reaction
$pp\to \pi^+pX$ at 1004, 1044 MeV, as new exotic baryon states is discussed.
The mass formula for the exotic baryons is constructed.
\end{minipage}
\end{center}

\vskip 10mm

\section{Supernarrow dibaryons}

In the present paper we will observe the works on a study of
supernarrow dibaryons (SNDs),
a decay of which into two nucleons is forbidden by the Pauli exclusion
principle \cite{fil1,fil2,alek,alek1}.
Such dibaryons satisfy the following condition:
\begin{equation}
(-1)^{T+S}P=+1
\end{equation}
where $T$ is the isospin, $S$ is the internal spin, and $P$ is the
dibaryon parity.
These dibaryons with the masses \mbox{$M < 2m_{N}+m_{\pi}$}
($m_N (m_{\pi}$) is the nucleon (pion) mass) can mainly decay
by emitting a photon. This is a new class of dibaryons
with the decay widths \mbox{$\le 1$keV}.
The experimental discovery of such states would have important consequences
for particle and nuclear physics and astrophysics.
In the following, we
summarize the experimental attempts to look for such dibaryons so far.

In ref. \cite{bilg} the existence of a dibaryon, called $d'$,
with quantum numbers $T=even$ and $J^P=0^-$, which forbid its decay into
two nucleons, with the mass $M=2.06$ GeV,
and the decay width $\Gamma_{\pi NN}=0.5$ MeV, has been
postulated to explain the observed resonance-like
behavior in the energy dependence of the pionic double charge exchange
on nuclei at an energy below the $\Delta$-resonance.
This dibaryon has a big mass and can decay into $\pi NN$.
However, there is a more conventional interpretation of these data.
It was shown \cite{nus} in the framework of the distorted-wave impulse
approximation that such a peak arises naturally because of the pion
propagation in the sequential process, in which pion double charge exchange
occurs through two successive $\pi N$ charge exchange reactions on two
neutrons.

In ref. \cite{khr} dibaryons with exotic quantum numbers were searched for
in the process $pp\to pp\gamma\gamma$. The experiment was performed with
a proton beam from the JINR Phasotron at an energy of about 216 MeV. The
energy spectrum of the photons emitted at $90^{\circ}$ was measured.
As a result, two peaks have been observed in this spectrum. This behavior
of the photon energy spectrum was interpreted as a signature of the exotic
dibaryon resonance with the mass of about 1956 MeV and possible isospin
$T=2$. In more details this experiment is considered in the reports of
Gerasimov and Khrykin.

On the other hand, an analysis \cite{cal} of the Uppsala proton-proton
bremsstrahlung data looking for the presence of a dibaryon in the mass range
from 1900 to 1960 MeV gave only the upper limits of 10 and 3 nb for
the dibaryon production cross section at proton beam energies of 200 and
310 MeV, respectively.
This result agrees with the estimates of the cross
section obtained at the conditions of this experiment in the framework
of the dibaryon production model suggested in ref. \cite{prc} and does
not contradict to the data of ref. \cite{khr}.

In ref. \cite{prc,konob,izv,yad,conf1, conf2, epj}, the reactions
$pd\to p+pX_1$ and
$pd\to p+dX_2$ were studied with the aim of searching for SND.
The experiment was
carried out at the Proton Linear Accelerator of INR with 305 MeV
proton beam using
the two-arm spectrometer TAMS. As was shown in ref. \cite{yad,prc},
the nucleons and the deuteron from the decay of SND into $\gamma NN$
and $\gamma d$ have to be emitted in a narrow angle cone with respect to
the direction of the dibaryon motion. On the other hand, if a dibaryon
decays mainly into two nucleons, then the expected  angular cone of
emitted nucleons must be more than $50^{\circ}$. Therefore, a detection
of the scattered proton in coincidence with the proton (or the deuteron)
from the decay of the dibaryon at correlated angles allowed the authors
to suppress essentially the contribution of the background processes and
to increase the relative contribution of a possible SND production.

Several software cuts have been applied to the mass spectra in these works.
In particular, the authors limited ourselves by the consideration of an
interval of the proton energy from the decay of the $pX_1$ states, which
was determined by the kinematics of the SND decay into $\gamma NN$ channel.
Such a cut is very important as it provides a possibility to suppress
essentially the
contribution from the background reactions and random coincidences.

In refs. \cite{conf1,conf2,epj}, CD$_2$ and $^{12}$C were
used as targets.
The scattered proton was detected in the left arm
of the spectrometer TAMS at the angle $\theta_L=70^{\circ}$. The second
charged particle (either $p$ or $d$) was detected in
the right arm by three telescopes located at $\theta_R=34^{\circ}$,
$36^{\circ}$, and $38^{\circ}$.

As a result,
three narrow peaks in the missing mass spectra have been observed  at
$M_{pX_1}=1904\pm 2$, $1926\pm 2$, and $1942\pm 2$ MeV  with widths
equal to the experimental
resolution ($\sim 5$ MeV) and with numbers of standard deviations (SD) of
6.0, 7.0, and 6.3, respectively. It should be noted
that the dibaryon peaks at $M_{pX_1}=1904$ and 1926 MeV had
been observed earlier by same authors in ref. \cite{prc,konob,izv,yad}
at somewhat different kinematical conditions.
The analysis of the angular distributions of the protons from the decay of
the $pX_1$ states showed that the peaks found can be explained as a
manifestation of the isovector SNDs, the decay of which into two nucleons
is forbidden by the Pauli exclusion principle.

An additional information about the nature of the observed states
was obtained by studying the missing mass $M_{X_1}$ spectra of the
reaction $pd\to p+pX_1$.
If the state found is a dibaryon decaying mainly into two nucleons then
$X_1$ is a neutron and the mass $M_{X_1}$ is equal to the neutron mass
$m_n$. If the value of $M_{X_1}$, obtained from the experiment, differs
essentially from $m_n$ then $X_1=\gamma+n$ and it is the additional
indication that the observed dibaryon is SND.

The simulation of the missing $M_{X_1}$ mass spectra for the reaction
$pd\to p+pX_1$ has been performed assuming that the SND decays as
SND$\to \gamma+^{31}S_0\to \gamma pn$ through two nucleon singlet state
$^{31}S_0$ \cite{fil2,prc,epj}. As a result, three narrow peaks at
$M_{X_1}=965$, 987, and 1003 MeV have been predicted. These peaks
correspond to the decay of the isovector SNDs with the masses 1904, 1926,
and 1942 MeV, respectively.

In the experimental missing mass $M_{X_1}$ spectrum
besides the peak at the neutron mass
caused by the process $pd\to p+pn$,
a resonance-like behavior of the spectrum has been observed at $966\pm 2$,
$986\pm 2$, and $1003\pm 2$ MeV. These values of $M_{X_1}$ coincide with
the ones obtained  from the simulation and differ essentially from
the value of the neutron mass (939.6 MeV). Hence, for all states under
study, we have $X_1=\gamma+n$ in support of a statement that the
dibaryons found are SNDs.

Recently, the reactions $pd\to pdX_2$ and $pd\to ppX_1$ have
been investigated by Tamii {\em et al.} \cite{tamii}
at the Research Center for Nuclear Physics
at the proton energy 295 MeV in the mass region of 1896--1914
MeV. They did not observe any narrow structure in this mass region
and obtained the upper limit of the production cross section of an
NN-decoupled dibaryon is equal to $\sim 2$ $\mu$b/sr if the dibaryon
decay width $\Gamma_D<< 1$ MeV. And if $\Gamma_D\simeq 3$ MeV, the
upper limit will be about $3.5\mu$b/sr. This limits are
smaller than the value of the cross section of $8\pm 4\mu$b/sr
declared in ref. \cite{prc}.

However, the latter value was overestimated that was caused by
not taking into account
angle fluctuations related to a beam position displacement on
the CD$_2$ target during the run.
As was shown in the next experimental runs, the real
value of the cross sections of the production of the SND with
the mass 1904 MeV must be smaller by 2--3 times than
that was estimated in \cite{prc}.

On the other hand, the simulation showed that the energy distribution
of the protons from the decay of the SND with the mass of 1904 MeV
has to be rather narrow with the maximum at $\sim 74$ MeV. This distribution
occupies the energy region of 60--90 MeV. In ref. \cite{tamii} the authors
considered the region 74--130 MeV. Moreover, they used a very large
acceptance of the spectrometer which detected these protons. As a result,
the ratio of the effect to the background in this work is more than 10
times worse than in ref. \cite{prc,epj}.
Very big errors and absence of a proper cut on the energy
of the protons from the decay of the $pX_1$ state in ref.
\cite{ tamii} did not allow the authors to observe any structure in the
$pX_1$ mass spectrum. For example, when the cut at $T_p=100$
was performed \cite{tamii2}, the obtained behavior
of this mass spectra would not contradict within the errors
the presence of the dibaryon peak in the considered mass region.

It is worth noting that the reaction $pd\to NX$ was investigated in
other works, too (see for example \cite{set}). However, in contrast to
the ref. \cite{prc,epj},
the authors of these works did not study
either the correlation between the parameters of the scattered proton and
the second
detected particle or the emission of the photon from the dibaryon decay.
Therefore, in these works the relative contribution of the dibaryons under
consideration was small, which hampered their observation.

\section{Exotic baryons}

As was shown above, in the missing $M_{X_1}$ mass spectra three peaks
at 966, 986, and 1003 MeV were observed.
On the other hand, the peak at $M_{X_1}=1003\pm $ MeV corresponds to
the resonance found in ref. \cite{tat} which was attributed to an exotic
baryon state $N^*$.

In ref. \cite{tat} Tatischeff {\em et al.} investigated
the reaction $pp\to\pi^+ X$ at energies
of $T_p=1520$, 1805, and 2100 MeV and at six angles, for each energy, from
$0^{\circ}$ up to $17^{\circ}$. Three peaks with widths about 5--8 MeV
have been observed in the missing
mass spectra of this reaction at $M_X=1004$, 1044, and 1094 MeV with a
statistic significance between 17 and 2 standard deviations. Two of these
masses are below the sum of the nucleon and pion masses.

If exotic baryons with anomalously small masses really exist,
the peaks observed at 966, 986, and 1003 MeV might be a manifestation of
such states. This is not in contradiction with the interpretation of the
peaks in the $M_{pX_1}$ mass spectra of ref. \cite{epj} as SNDs, since, in
principal, SND could decay into $NN^*$. In this case the SND decay width
could be equal to a few MeV.

The existence of such exotic states, if
proved to be true, will fundamentally change our understanding of the
quark structure of hadrons.

Exotic baryon states with masses smaller than $m_N+m_{\pi}$ can decay mainly
with an emission of photons. If they decay into $\gamma N$ then such states
have
to contribute to the Compton scattering on the nucleon. However, L'vov and
Workman \cite{lvov} showed that existing experimental data on this process
"completely exclude" such exotic baryons as intermediate states in the Compton
scattering on the proton. On the other hand, the early Compton scattering
data were not accurate enough to rule out these baryon resonances. Moreover,
a measurement of the process $\gamma p\to\gamma p$ in the photon energy range
$60< E_{\gamma}<160$ MeV resulted in a peak at $M\approx 1048$ MeV
with an experimental resolution of 5 MeV and with 3.5 standard deviations
\cite{beck}. Unfortunately, the accuracy of this experiment is not enough
to do an unambiguous conclusion about the $N^*$ contribution to the Compton
scattering on the nucleon.

In Ref. \cite{kob} it was assumed that these states could belong to the
totally antisymmetric $\underline{20}$-plet of the spin-flavor SU(6)$_{FS}$
symmetry. Such a $N^*$ can transit into a nucleon only if two quarks from
the $N^*$ participate in the interaction. Then the simplest decay of the
exotic baryons with the small masses is $N^*\to \gamma\gamma N$.

On the other hand, the $N^*$s were produced in ref. \cite{tat,epj}, more
probably, from the decay of 6-quark states,
what is supported by the observation
of the dibaryon resonances in \cite{epj}. Therefore, an exotic quark
structure of the $N^*$ could be arisen which suppressed, in particular, 
the decay $N^*\to \gamma N$ and could be the reason of an unobservation of 
such states early.
In order to clarify the question about an existence of such exotic baryons,
different experiments were proposed, in particular, in ref. \cite{mainz}.
In the present work we will assume that such states exist.

In ref. \cite{tat} it was shown that values of the masses of the
baryon resonances, observed in this work, can be reproduced with good
enough accuracy by the mass formula
for two colored clusters of quarks at the end of a stretched bag which was
derived in terms of color magnetic interactions \cite{muld,besl}.
There are two free parameters in this model which were fixed by requiring
the mass of the nucleon and that of the Roper resonance to be reproduced
exactly. As a result of the calculations, the following values of the
masses and possible isospin $(I)$ and spin $(J)$ of these
baryons have been obtained:
$$ 
 M(I;J)=1005(1/2;1/2,3/2),\qquad 1039(1/2;3/2).
$$

N. Konno \cite{kon} pointed out that the masses of the exotic baryons from
the ref. \cite{tat} can also be reproduced by the formula of the diquark
cluster model \cite{kon2}. Eight free parameters of this model were fixed
using data of baryon masses and the $\pi d$ phase shift.
This model predicted the following values of the masses and $I,J^P$:
$$
M=990(I=1/2(J^P=1/2^-); 3/2(1/2^-),
$$
$$
M=1050(1/2(1/2^-,3/2^-); 3/2(1/2^-,3/2^-)), \qquad M=1060(1/2(1/2^-,3/2^-)).
$$

However, these two models do not reproduce the values of masses: 966 and
986 MeV, obtained in \cite{epj}.

Th. Walcher \cite{walch} noted that the masses taking all experiments
together and
including the neutron ground state and two additional masses at 1023 and
1069 MeV are equidistant within the errors with an average mass difference
of $\Delta M=21.2\pm 2.6$ MeV. The author hypothesized the existence of a
light Goldstone boson with the mass of 21 MeV consisting of light current
quarks. It was assumed that the series of excited states is
due to the nucleon in its ground state plus 1, 2, 3,... light Goldstone
bosons as the quantum of excitation.

In the present paper we construct a model which allows us to calculate
the masses of all possible exotic baryon states below the $\pi$ meson
production threshold and determine their parities.
This model is based on the
calculation of the contribution of meson--baryon loops to the exotic
baryon mass operator.

An analysis of the mass shifts of experimentally well-known baryons due to
meson-baryon loops was carried out in a set of works (see for references
\cite{cap}). In these works, the self energy of a baryon was
calculated, as a rule, in the framework of a time-ordered perturbation
theory. In this case, an underintegral expression diverges strongly and
additional assumptions about a behavior of baryon-baryon-meson vertices
are required.

We will calculate the masses of exotic baryons with
help of dispersion relations with two subtractions for the mass operator.
The mass operator is determined as
\begin{equation}
\hat p-M=\hat p-m-\Sigma(M), \qquad \Sigma=a\hat p+b
\end{equation}
where $M$ and $p$ are the mass
and the 4-momentum of the $N^*$ under consideration, $m$
is the mass of the baryon in the intermediate state.
The mass of the $N^*$ is equal to
\begin{equation}
\label{M}
M=m+\delta
\end{equation}
where
\begin{equation}
\delta=\bar u(p)\Sigma u(p)=\bar u(p)(aM+b)u(p).
\end{equation}

In order to find $\delta$ we construct the dispersion relations over
$M^2$ for $\delta(M)$ with two subtraction at $M^2=m^2$. Then taking into
account eq.(\ref{M}) we obtain the following nonlinear integral equation for
the mass $M$
\begin{equation} 
\label{ds}
M=m+Re\,\delta(m)+\left.(M^2-m^2)\frac{d\,Re\,\delta(M)}{d\, M^2}
\right|_{M=m}+
\frac{(M^2-m^2)^2}{\pi}P\int\limits_{(m+\mu)^2}^{\infty}
\frac{Im\,\delta(x)\,dx}{(x-M^2)(x-m^2)^2}.
\end{equation}   

We will consider here only baryon with the spin equal to 1/2.
Then the function $Im\,\delta$ for the baryon-pion loop can be written as
\begin{equation}  
\bar u(p)Im\,\delta(M)u(p)=\bar u(p)Im\,\Sigma(M)u(p)=
N\bar u(p)(\mp\hat p_1+m)u(p)
\end{equation} 
where $N=1/2(g^2/4\pi)|p_1|/M$,
$p_1$ is the 4-momentum of the baryon in the intermediate state.
The sign minus (plus) at $\hat p_1$
corresponds to the same (opposed) parities of the final baryon and the
baryon in the intermediate state.
Multiplying these expressions by $u(p)$ from the left and by
$\bar u(p)$ from right and using the condition
$u(p)\bar u(p)=(\hat p+M)/2M$ we have
\begin{equation}
(\hat p+M)Im\,\delta(M) (\hat p+M)=N(\hat p+M)(\mp\hat p_1+m)(\hat p+M).
\end{equation}
Calculating traces in the left and the right parts of this expression
we obtain:
\begin{equation}
Im\,\delta(M)=\frac{N}{2M^2}[\mp 2(pp_1)M+(p^2+M^2)m].
\end{equation}
Taking into account that $p^2=M^2$ , $2(pp_1)=M^2+m^2-\mu^2=2M E(M)$ we
have
\begin{equation}
Im\,\delta(M)=\frac12\frac{g^2}{4\pi}\frac{|p_1|}{M}(m\mp E),
\end{equation}
here $|p_1|=\sqrt{E^2-m^2}$.

Two subtractions in the dispersion relations provide a very good convergence
of the underintegral expression in eq.(\ref{ds}). Therefore, we restrict
ourselves to a consideration only one baryon and the pion in the
intermediate state. The calculations showed that the
contribution of the $\sigma$ meson is negligible. Therefore, it is expected
that the contribution of other heavy mesons
in the mass region under consideration is negligible too. However,
these contributions could be important in the mass region higher than
the $\pi N$ production threshold.

As the subtraction is performed at the mass shell of the baryon in
the intermediate state,
the subtraction constant $Re\,\delta(m)$ is equal to zero. We
assume also that $dRe\,\delta(M)/dM^2|_m=0$. This assumption corresponds to
a supposition that the baryon with the mass $m$ is in the ground state.
It should be noted that if one takes
into account a few different baryons in the intermediate state, the
subtraction constants could not be equal to zero because
in this case the mass $m$ does not
coincide with the mass shell for some of these baryons.

In our calculations we
assumed that all baryons under consideration have the isotopic
spin and the spin equal to $1/2$. Then eq.(\ref{ds}) has a solution
in the mass region lower than the $\pi N$ production threshold only if
the final baryon with the mass $M$ and the baryon in the intermediate
state have opposite parities.

Taking the nucleon plus the pion ($\pi^+$ and $\pi^0$) in the
intermediate state, we find for the first exotic baryon state
the mass $M=963.4$ MeV and
$J^P=1/2^-$. Then taking the first exotic baryon with $m=963.4$ MeV,
$J^P=1/2^-$ and the $\pi$ meson in the intermediate state
we obtain for the second exotic baryon state $M=987.0$ MeV and
$J^P=1/2^+$.
Continuing the same procedure, six possible exotic states of baryons
have been found. The obtained states are listed in Table 1 where the
experimental data are also given.
\begin{table}
\centering
\caption{The masses and $J^P$ of the exotic baryon resonances}
\label{table1}
\begin{tabular}{|c|c|c|c|c|}\hline
     &          &  model   &  experiment & experimental \\
$N^*$& $J^P$    &$M$ (MeV) & $M$ (MeV)   & works        \\ \hline
 1   &$\frac12^-$& 963.4   & $966\pm 2$  & \cite{epj} \\ \hline
 2   &$\frac12^+$& 987   & $986\pm 2$  & \cite{epj} \\ \hline
 3   &$\frac12^-$& 1010  & $1004\pm 2$ &\cite{tat,epj}\\ \hline
 4   &$\frac12^+$& 1033    &    ?        &               \\ \hline
 5   &$\frac12^-$& 1056    & $1044\pm 2$ & \cite{tat}    \\ \hline
 6   &$\frac12^+$& 1079    &    ?        &               \\ \hline
\end{tabular}
\end{table}

As seen from this table, the results of calculations are in a good
agreement with the experimental data. The mass values of two unobserved
still states at 1033 and 1079 MeV are close to the ones predicted in
\cite{walch}.

At the calculation we assumed that the coupling constant
$g^2/4\pi$ in the vertices $NN^*\pi$ and $N^*_iN^*_f\pi$
is same for all $N^*$ and
equal to the coupling constant of the $\pi NN$ interaction
($g^2_{pp\pi^0}/4\pi=14.6$). So, an increase of the difference between
the model predictions for the masses and experimental data
with a rise of the mass could be caused, in
particular, by this assumption. Therefore, it is expected that a real
value of the mass for the unobserved still state \#6 is smaller by
$\sim 10$ MeV and so it has to be bellow the pion production threshold.


As follows from the table, the odd parity has been predicted for
all baryon states observed in \cite{tat}.
It agrees with the comment of Th. Walcher \cite{walch}
and is due to the kinematics of the experiment \cite{tat}. This experiment
detected $\pi^+$ and $p$ from the reaction $pp\to\pi^++pX$ in coincidence
in one spectrometer at small angles. Since the $N^*$ states
were observed mainly in the forward direction with respect to the beam
($\theta\approx 5^{\circ}$), this means that the outgoing $\pi$ and $p$
carry no angular momentum and only the odd parity state of $N^*$ can be
observed in this experiment. But the baryons with the odd parity
cannot belong to the total antisymmetric $\underline{20}$-plet \cite{kob}.

It is worth noting that eq.(\ref{ds}) has additionally solutions. But all
of them are essentially higher than $\pi N$ production threshold.
However, to get more reliable information about values of the masses
obtained in this mass region, we should take into account the contribution
of the $\sigma$, $\omega$, and other mesons.

\section{Conclusion}

As a result of the study of the reaction $pd\to p+pX_1$,  three narrow
peaks at $M_{pX_1}=1904$, 1926, and 1942 MeV have been observed. The analysis
of the angular distributions of the protons from the decay of the $pX_1$
states showed that the peaks found can be explained as a manifestation of the
isovector SNDs, the decay of which into two nucleons is forbidden by the
Pauli exclusion principle. The observation of the peaks in the missing mass
$M_{X_1}$ spectra at 966, 986, and 1003 MeV is an additional indication
that the dibaryons found are the SNDs.

On the other hand, these peaks in $M_{X_1}$ mass spectra and peaks observed
in \cite{tat} in the reaction $pp\to\pi^+pX$
could be consider as the new exotic baryon states with small
masses. However, additional experiments are necessary to understand the
real nature of these peaks.
In the present paper, the mass equation has been constructed which was used
to calculate the masses and determine parities of the exotic baryons.
The obtained values of the masses are in a good agreement with the
experimental data \cite{epj,tat}. Two new exotic baryon
states bellow the $\pi$ production threshold have been predicted.

The author thanks R. Beck, V. Fainberg, A. Kobushkin, B. Tatischeff,
and Th. Walcher for helpful discussions.

\end{document}